\documentclass[journal,twoside,web]{ieeecolor2}
\usepackage{generic}
\usepackage{cite}
\usepackage{amsmath,amssymb,amsfonts}
\usepackage{algorithmic}
\usepackage{graphicx}
\usepackage{graphics}
\usepackage{makecell}
\usepackage{textcomp}
\usepackage[normalem]{ulem}
\def\BibTeX{{\rm B\kern-.05em{\sc i\kern-.025em b}\kern-.08em
    T\kern-.1667em\lower.7ex\hbox{E}\kern-.125emX}}
\usepackage{subfigure}
\usepackage{tabularx}
\usepackage{multirow}

\DeclareMathOperator*{\argmax}{argmax\;}

\newcommand{\squotes}[1]{`#1'}  
 
\newcommand{\etal}{et al.\ }
\newcommand{\sref}[1]{Section~\ref{#1}}
\newcommand{\fref}[1]{Figure~\ref{#1}}
\newcommand{\tref}[1]{Table~\ref{#1}}
\newcommand{\eref}[1]{(\ref{#1})}
\newcommand{\code}{\texttt}
\definecolor{red}{RGB}{255, 0, 0}

\begin{document}
\title{AADNet: An End-to-End Deep Learning Model for Auditory Attention Decoding}
\author{Nhan Duc Thanh Nguyen, Huy Phan, Simon Geirnaert, Kaare Mikkelsen, and Preben Kidmose, \IEEEmembership{Member, IEEE}
\thanks{This work was funded by the William Demant Foundation, grant numbers 20-2673, and supported by Center for Ear-EEG, Department of Electrical and Computer Engineering, Aarhus University, Denmark.}
\thanks{Nhan Duc Thanh Nguyen (corresponding author, e-mail: {\tt\small ndtn@ece.au.dk}), Kaare Mikkelsen (e-mail: {\tt\small mikkelsen.kaare@ece.au.dk}), and Preben Kidmose (e-mail: {\tt\small pki@ece.au.dk}) are with the Center for Ear-EEG, Department of Electrical and Computer Engineering, Aarhus University, 8200 Aarhus N, Denmark.}
\thanks{Huy Phan (e-mail: {\tt\small huyphan@meta.com}) is with Meta Reality Labs, Paris 75002, France. The work does not relate to H.P.'s work at Meta.}
\thanks{Simon Geirnaert ({\tt\small simon.geirnaert@esat.kuleuven.be}) is with the Department of Electrical Engineering (ESAT), Stadius Center for Dynamical Systems, Signal Processing and Data Analytics, KU
Leuven and Leuven.AI - KU Leuven institute for AI, Kasteelpark Arenberg 10, B-3001 Leuven, Belgium, and the Department of Neurosciences, Research Group ExpORL, Herestraat 49 box 721, B-3000 Leuven, Belgium. His research is supported by a junior postdoctoral fellowship fundamental research from the Research Foundation Flanders (FWO) (1242524N).}}

\maketitle

\begin{abstract}
Auditory attention decoding (AAD) is the process of identifying the attended speech in a multi-talker environment using brain signals, typically recorded through electroencephalography (EEG). Over the past decade, AAD has undergone continuous development, driven by its promising application in neuro-steered hearing devices. Most AAD algorithms are relying on the increase in neural entrainment to the envelope of attended speech, as compared to unattended speech, typically using a two-step approach. First, the algorithm predicts representations of the attended speech signal envelopes; second, it identifies the attended speech by finding the highest correlation between the predictions and the representations of the actual speech signals. In this study, we proposed a novel end-to-end neural network architecture, named AADNet, which combines these two stages into a direct approach to address the AAD problem. We compare the proposed network against traditional stimulus decoding-based approaches, including linear stimulus reconstruction, canonical correlation analysis, and an alternative non-linear stimulus reconstruction using three different datasets. AADNet shows a significant performance improvement for both subject-specific and subject-independent models. Notably, the average subject-independent classification accuracies for different analysis window lengths range from 56.3\% (1 s) to 78.1\% (20 s), 57.5\% (1 s) to 89.4\% (40 s), and 56.0\% (1 s) to 82.6\% (40 s) for three validated datasets, respectively, showing a significantly improved ability to generalize to data from unseen subjects. These results highlight the potential of deep learning models for advancing AAD, with promising implications for future hearing aids, assistive devices, and clinical assessments.
\end{abstract}


\section{Introduction}\label{s_introduction}
\IEEEPARstart{T}{he} human brain demonstrates a remarkable ability to segregate sound streams, allowing individuals to focus on the sound of interest while disregarding the others. For individuals with hearing impairment, this ability is often significantly deteriorated \cite{shinn-cunninghamSelectiveAttentionNormal2008}. Assistive hearing devices equipped with noise suppression and speech enhancement algorithms can partially compensate for this deficit. However, these hearing devices tend to perform poorly in a multi-talker environment due to a lack of information about the target talker. Therefore, over the past decade, various research has been conducted to address this challenge by identifying the attended talker from brain signals, i.e., auditory attention decoding (AAD). This can potentially lead to the development of so-called neuro-steered hearing devices, in which the attended speech is enhanced by suppressing irrelevant sound sources. While this paper focuses specifically on the AAD challenge itself, alternative approaches in neuro-steered hearing devices address this issue by simultaneously performing speech separation and enhancement informed by brain signals, known as brain-informed speech separation \cite{ceoliniBraininformedSpeechSeparation2020,zhangBASENTimeDomainBrainAssisted2023}.

%Over the past decades, various studies have been conducted to address this challenge, leading to the development of a new generation of hearing devices known as neuro-steered assistive hearing devices. One potential approach is to improve speech separation and enhance the target speech using information decoded from brain signals, i.e. brain-informed speech separation \cite{ceoliniBraininformedSpeechSeparation2020, zhangBASENTimeDomainBrainAssisted2023}. Alternatively, another approach -- identifying the attended talker, i.e. auditory attention decoding (AAD) -- has garnered significant attention from researchers, as it can also help enhance target speech by suppressing untargeted sound sources.}

Previous studies have demonstrated neural encoding of various speech features using different recording techniques, including the speech envelope via EEG \cite{aikenHumanCorticalResponses2008}, spectrogram via electrocorticography (ECoG) \cite{mesgaraniSelectiveCorticalRepresentation2012}, and via magnetoencephalography (MEG) \cite{dingEmergenceNeuralEncoding2012, dingNeuralCodingContinuous2012, akramRobustDecodingSelective2016}. As EEG is minimally invasive and can be integrated into portable devices, following the study of Aiken \etal \cite{aikenHumanCorticalResponses2008}, the majority of AAD methods are based on the envelope following response extracted from EEG signals, and these methods have successfully demonstrated AAD across various experimental paradigms \cite{hortonEnvelopeResponsesSingletrial2014a, osullivanAttentionalSelectionCocktail2015, mirkovicDecodingAttendedSpeech2015, fuglsangNoiserobustCorticalTracking2017, biesmansAuditoryInspiredSpeechEnvelope2017b, decheveigneDecodingAuditoryBrain2018a, detaillezMachineLearningDecoding2020, ciccarelliComparisonTwoTalkerAttention2019a, thorntonRobustDecodingSpeech2022}. This approach has become a dominant and well-established method for addressing the AAD problem. While the aforementioned methods rely on acoustical features, another approach has investigated linguistic and lexical features, such as the onset and surprisal of words and phonemes, to decode neural responses to speech \cite{puffayRobustNeuralTracking2023}. These approaches assume that words and phonemes are annotated and that language-specific dictionaries with word and phoneme probabilities are accessible. Additionally, Raghavan \etal \cite{raghavanImprovingAuditoryAttention2024} proposed a system that identifies auditory events using the local maxima in the envelope rate of change, and utilized masking-specific event-related potential classifiers to determine the attended sound source, suggesting a new approach to AAD.

The most common approach used in envelope-based AAD algorithms is backward modeling in which a decoder is trained to reconstruct the attended speech envelope. During the training procedure, the Pearson correlation or the mean square error between the reconstructed and actual envelopes is used as the objective function to optimize the model's parameters. The reconstructed envelope is then correlated with the actual envelopes and used as input to an additional classifier to determine the attended speech stream. Many studies have used linear decoders across various recording paradigms \cite{hortonEnvelopeResponsesSingletrial2014a, osullivanAttentionalSelectionCocktail2015, mirkovicDecodingAttendedSpeech2015, fuglsangNoiserobustCorticalTracking2017}, while others have employed non-linear models with different neural network structures, such as fully connected neural networks (FCNN) \cite{detaillezMachineLearningDecoding2020, thorntonRobustDecodingSpeech2022}, convolutional neural networks (CNN) \cite{thorntonRobustDecodingSpeech2022}, long-short term memory (LSTM) \cite{xuAuditoryAttentionDecoding2022a}, and CNN-LSTM \cite{kuruvilaExtractingAuditoryAttention2021}, achieving promising results. Generally, the linear models seem to be consistent and well-established due to their simplicity (a low number of parameters) and have been applied across various datasets. The non-linear counterparts typically have large numbers of parameters and have been reported to outperform the linear models \cite{thorntonRobustDecodingSpeech2022, xuAuditoryAttentionDecoding2022a} due to their capacity to model the nonlinearity of speech processing in the auditory system. However, most of these non-linear methods are designed and validated on specific paradigms from the original studies, and they may not generalize well to other datasets or recording paradigms.

An alternative approach to backward modeling is forward modeling, which predicts the EEG response from the speech envelopes. The predicted EEG is then used to compare with the measured EEG to determine the attended speech. This approach has been reported to underperform compared to the backward models \cite{wongComparisonRegularizationMethods2018a}. This result appears reasonable, given that neural responses arise from a large number of underlying neural sources, of which only a small subset are directly or indirectly related to the audio. Other studies have attempted a combined forward-backward approach using the Canonical Component Analysis (CCA) algorithm \cite{decheveigneDecodingAuditoryBrain2018a, alickovicTutorialAuditoryAttention2019b} to convert both EEG and envelope signals into maximized correlated latent vectors, followed by a classifier to determine the attended speech. This approach has been demonstrated to outperform the other linear decoder methods \cite{geirnaertElectroencephalographyBasedAuditoryAttention2021}.

Common to the approaches mentioned above is that they involve two separate stages: training models to convert the EEG signals and/or the audio envelopes into latent vectors, calculating the correlation scores, and using a classifier to decode the attention. Recently, Ciccarelli \etal \cite{ciccarelliComparisonTwoTalkerAttention2019a} have proposed a direct approach using a CNN model that does not explicitly reconstruct the envelope. This new approach has been shown to outperform the previous linear and non-linear envelope reconstruction approaches. However, like the majority of other AAD studies, this approach has not been tested for subject-independent models, which is an important yet underdeveloped topic, as pointed out by other authors \cite{puffayRelatingEEGContinuous2023}.

The motivation for this study is to improve AAD performance, specifically, its generalization capability to unseen subjects, taking a step towards integrating AAD algorithms into real-life applications. To achieve this, we adopt the direct approach by proposing AADNet, an end-to-end deep learning (DL) model to address the AAD problem. The model utilizes EEG signals and the audio envelopes of two speakers to directly determine the attended speaker without reconstructing the attended envelopes. To evaluate the performance of the proposed model, we conduct a comparison with other state-of-the-art AAD methods: linear stimulus reconstruction (LSR) \cite{osullivanAttentionalSelectionCocktail2015}, CCA \cite{ decheveigneDecodingAuditoryBrain2018a}, and non-linear stimulus reconstruction (NSR) \cite{thorntonRobustDecodingSpeech2022} on three datasets for both subject-specific (SS) and subject-independent (SI) models. The results demonstrate that the proposed model outperforms the other methods, particularly in the subject-independent setting. The code is publicly available at https://github.com/babibo180918/AADNet.

\section{Envelope-based AAD algorithms}\label{s_algs}
As previously mentioned, current AAD algorithms predominantly focus on the correlation between the recorded EEG signals and the envelope of the attended stimulus. According to Geirnaert \etal \cite{geirnaertElectroencephalographyBasedAuditoryAttention2021}, the most robust AAD methods are the LSR and the CCA model that combines the forward and backward approaches. Therefore, in this study, we implemented these methods as least-squares-based baselines to compare with the proposed method. Apart from that, with breakthroughs across multiple disciplines, deep learning (DL) methods have been introduced in various studies to advance the field \cite{detaillezMachineLearningDecoding2020,ciccarelliComparisonTwoTalkerAttention2019a,monesiLSTMBasedArchitecture2020, kuruvilaExtractingAuditoryAttention2021, thorntonRobustDecodingSpeech2022, xuAuditoryAttentionDecoding2022a,puffayRobustNeuralTracking2023,accouDecodingSpeechEnvelope2023}. However, these methods address different tasks (single-speaker match-mismatch, AAD speaker identification (SpkI), and locus of attention (LoA)), utilize different recording modalities (EEG, MEG), and employ various speech features (linguistics, speech envelope, spectrogram). A direct comparison of all these methods is beyond the scope of this study. In this study, we include DL-based studies that meet the following criteria:
\begin{itemize}
  \item Recent DL-based model, where the original paper proposing the model did not utilize both the two publicly available datasets DTU and KUL (see \sref{sss_dataset_2} and \sref{sss_dataset_3}).
  \item Methods designed for the AAD speaker identification task using envelope features.
  \item Methods with publicly available source code. 
\end{itemize}
Typically, other studies use specific datasets to develop and tune their proposed models. This study follows a similar approach by using the EventAAD dataset (see \sref{sss_dataset_1}) to develop our model, and only using the DTU and KUL datasets for validating the model. Therefore, the first criterion ensures that the hyperparameters of the included models were not tailored to the datasets used in the model comparison. Based on these criteria, the study by Thornton \etal \cite{thorntonRobustDecodingSpeech2022} is considered the state-of-the-art for NSR methods.

\begin{figure}[!ht]
\centering
\includegraphics[width=1.0\linewidth]{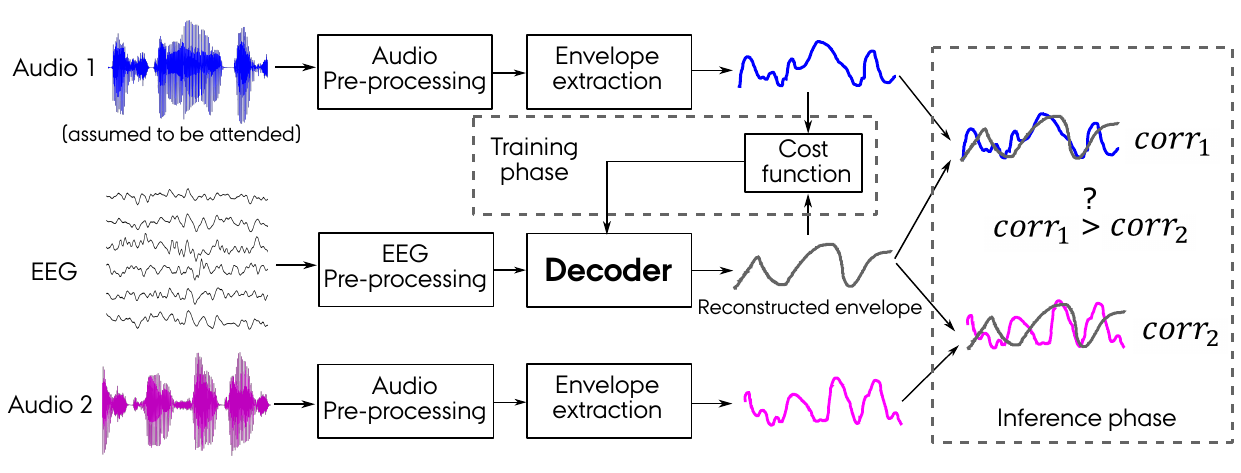}
\caption{Schematic depiction of stimulus reconstruction-based AAD.}
\label{fig_SR_AAD}
\end{figure}

The concept of decoding attention based on a stimulus reconstruction approach is depicted in \fref{fig_SR_AAD}. First, multi-channel EEG signals and audio signals from each stream are pre-processed (see \sref{ss_preprocess}). Subsequently, the actual envelopes of the audio signals in each audio stream are extracted, assuming the demixed speech streams are available. In the training phase, the envelope of the attended stream and the EEG signals are used to train a decoder. In the inference phase, the EEG signals are used to reconstruct the envelope of the attended stream. The attended stream is determined as the one whose envelope has the highest correlation with the reconstructed envelope. The decoder can be constructed as a linear or non-linear model that maps the EEG data to the attended or unattended audio envelope. However, as found by O'Sullivan \etal \cite{osullivanAttentionalSelectionCocktail2015}, the attended decoder obtains higher decoding accuracy than the unattended one does. Therefore, in this study, we use the attended decoder.

\subsection{Linear methods}\label{ss_linear}
\subsubsection{Linear stimulus reconstruction}\label{sss_SR_AAD}
We implemented the linear stimulus reconstruction (LSR) model, introduced by O'Sullivan \etal \cite{osullivanAttentionalSelectionCocktail2015}. For the set of $N$ electrodes, the decoder, constructed as a spatiotemporal filter, maps the recorded EEG response to the stimulus envelope as follows:

\begin{flalign}
\hat{s}(t) &= \sum\limits_{n=1}^{N}\sum\limits_{\tau=0}^{L-1}g(\tau,n)x(t+\tau, n) \label{eq.1},
\end{flalign}
where $\hat{s}(t)$ is the reconstructed envelope signals at time $t$, $x(t+\tau, n)$ is the recorded EEG signal at time $(t+\tau)$ from electrode $n$, $\tau$ is the time lag index, ranging from $0$ to $L-1$, and $g(\tau,n)$ is the coefficient of the decoder at electrode $n$ and time lag $\tau$. It is noted that the decoder is anticausal because the audio stimulus causes the EEG response. The decoder was estimated to minimize the mean squared error between the original and the reconstructed envelopes. To prevent overfitting, the ridge regularization method \cite{wongComparisonRegularizationMethods2018a} is used, leading to the following solution:

\begin{flalign}
\mathbf{g} &= (\mathbf{X}^{\textsc{t}}\mathbf{X} + \mathbf{I}\lambda)^{-1}\mathbf{X}^{\textsc{t}}{\mathbf{s}}, \label{eq.2}
\end{flalign}
where $\mathbf{g} \in \mathbb{R}^{NL\times 1}$  is the decoder, collecting all
decoder coefficients for all time lags and channels. Assuming that there are $T$ envelope samples available, $\mathbf{s} \in \mathbb{R}^{T\times 1}$ is the envelope vector, $\mathbf{X} \in \mathbb{R}^{T\times NL}$ is the EEG matrix with each row vector contains all the EEG signals across all time lags and channels, $\mathbf{I}$ is the identity matrix, and $\lambda$ is the regularization parameter, which is estimated from a cross-validation approach from a set of values ranging from $10^{-2}$ to $10^{10}$ with a logarithmic step. The time lag $\tau$ covers a temporal EEG context from 0 up to 250$\,$ms post-stimulus, as it has been found to have the best decoding accuracy \cite{osullivanAttentionalSelectionCocktail2015}.

\subsubsection{CCA}\label{sss_CCA_AAD}
CCA is a multivariate statistical technique used to analyze the relationship between two sets of variables \cite{hotellingRelationsTwoSets1936}. The CCA method has been successfully applied to the AAD problem and has achieved promising results \cite{decheveigneDecodingAuditoryBrain2018a, alickovicTutorialAuditoryAttention2019b}. In AAD, the goal of CCA is to find an optimal spatiotemporal linear transform (decoder) $\mathbf{w}_x \in \mathbb{R}^{NL\times 1}$ on EEG signals $\mathbf{X} \in \mathbb{R}^{T\times{NL}}$ and a temporal linear transform (encoder) $\mathbf{w}_s \in \mathbb{R}^{L_{s}\times 1}$ on audio envelopes $\mathbf{S} \in \mathbb{R}^{T\times L_{s}}$ to maximize the correlation between the two latent vectors. $L$ and $L_s$ correspond to the decoder length and encoder length, respectively. The CCA method can be described as the following optimization equation:

\begin{flalign}
\mathbf{\hat{w}}_x, \mathbf{\hat{w}}_s &= \argmax_{\mathbf{w}_x, \mathbf{w}_s} {\frac{\mathbf{w}_x^{\textsc{t}}\mathbf{R_{xs}}\mathbf{w}_s}{\sqrt{\mathbf{w}_x^{\textsc{t}}\mathbf{R_{xx}}\mathbf{w}_x}\sqrt{\mathbf{w}_s^{\textsc{t}}\mathbf{R_{ss}}\mathbf{w}_s}}},\label{eq.4}
\end{flalign}
where $\mathbf{R_{xx}} \in \mathbb{R}^{NL\times{NL}}$, $\mathbf{R_{ss}} \in \mathbb{R}^{L_{s}\times{L_{s}}}$, and $\mathbf{R_{xs}} \in \mathbb{R}^{NL\times{L_{s}}}$ are the autocorrelation matrix of $\mathbf{X}$, autocorrelation matrix of $\mathbf{S}$, and cross-correlation matrix of $\mathbf{X}$ and $\mathbf{S}$, respectively.

By solving a generalized eigenvalue decomposition, the solution for \eref{eq.4} can be retrieved with a pair of decoders corresponding to the largest eigenvalue. The solution can be extended to a set of $J$ pairs of decoders ($\mathbf{w}_x \in \mathbb{R}^{NL\times J}$) and encoders ($\mathbf{w}_s \in \mathbb{R}^{L_{s}\times J}$) corresponding to $J$ CCA components, $J = min(L, L_s)$. $J$ Pearson correlation coefficients between the outputs $J$ decoders and encoders can be retrieved accordingly. To determine the attended speaker, in this study, we used a linear discriminant analysis (LDA) classifier, which is recommended in literature \cite{geirnaertElectroencephalographyBasedAuditoryAttention2021, alickovicTutorialAuditoryAttention2019b}, taking the differences of $J$ Pearson correlation coefficients of both competing speakers as the input. The encoder length $L_s$ and decoder length $L$ were set to $1.25\,s$ (pre-stimulus lags) and $250\,ms$ (post-stimulus lags), respectively, according to the optimal values used in \cite{geirnaertElectroencephalographyBasedAuditoryAttention2021}. The method for determining the $J$ value is described in \sref{ss_implementations}.

\subsection{Nonlinear methods}\label{ss_nonlinear}

\subsubsection{Non-linear stimulus reconstruction}
Another approach is reconstructing the stimulus envelope using a non-linear model (NSR). In this study, we implemented the CNN-based network proposed by Thornton \etal \cite{thorntonRobustDecodingSpeech2022}, which meets the criteria described above, as a baseline method for the NSR approach. The network was inspired by the EEGNet architecture developed by Lawhern \etal \cite{lawhernEEGNetCompactConvolutional2018}, which comprises two main convolutional blocks, one fully connected (FC) classification layer and employs the exponential linear unit (ELU) as a nonlinear activation function, as well as batch normalization (BN) and average pooling. Details of the network can be found in the original study \cite{thorntonRobustDecodingSpeech2022}. Another notable study using the NSR approach is by De Taillez \etal \cite{detaillezMachineLearningDecoding2020}. The authors developed a feed-forward neural network comprising a single hidden layer and an output layer with other DL features such as \squotes{tanh} activation functions, dropout (DO) \cite{srivastavaDropoutSimpleWay2014}, and BN. Despite our best efforts to implement and validate this method, its performance was significantly lower compared to the other methods, and therefore, the results are not included here. A similar observation was reported in \cite{geirnaertElectroencephalographyBasedAuditoryAttention2021}.

\subsubsection{Proposed direct AAD method}\label{sss_proposed_AAD}

\text{\\}

\begin{figure}[!ht]
\centering
\includegraphics[width=1.0\linewidth]{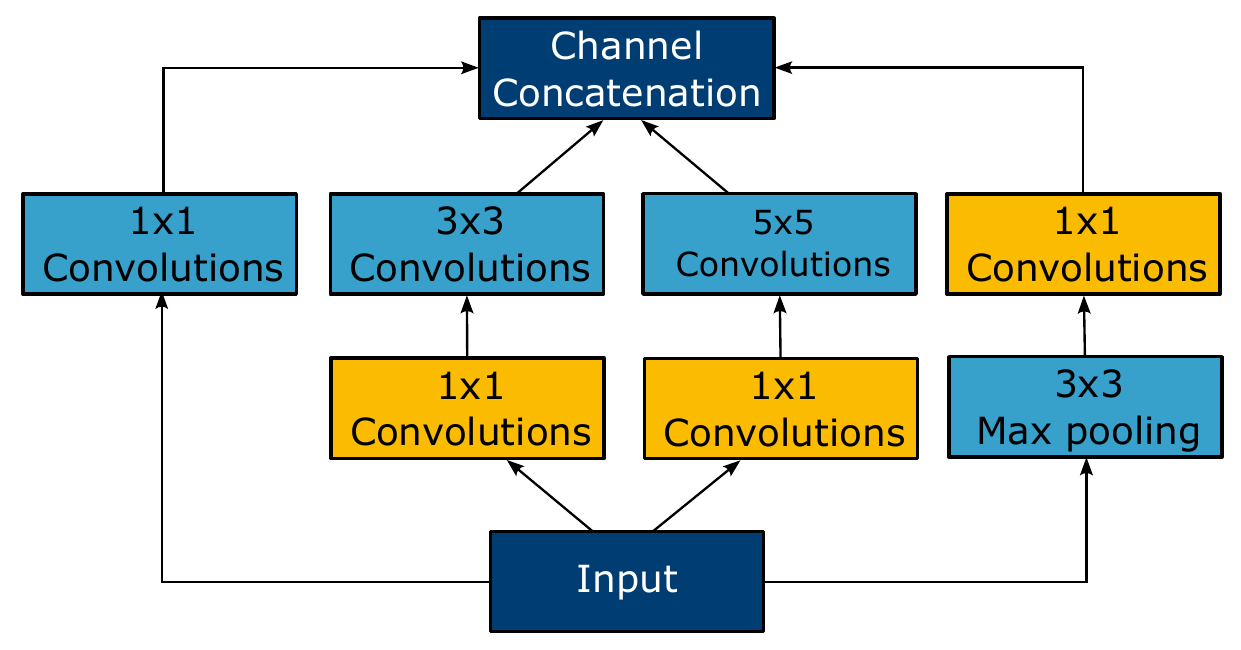}
\caption{Structure of the Inception block from the original study \cite{szegedyGoingDeeperConvolutions2015} with 2D convolutions. It is default that there is a ReLU activation function in each branch before the outputs are concatenated.}
\label{fig_Inception}
\end{figure}

\textbf{Inception backbone: }
Inception is a basic convolutional block, proposed in GoogLeNet \cite{szegedyGoingDeeperConvolutions2015} to solve a problem of object detection and image classification. \fref{fig_Inception} depicts the structure of an Inception block. It consists of four parallel branches. The first three branches are the convolutional layers with kernel sizes of $1\times1$, $3\times3$, and $5\times5$ respectively. The $1\times1$ convolution in the first branch transforms the features from the earlier layers to the later layers (if the network contains multiple Inception blocks stacked on top of each other) while the other convolutions in the middle branches extract spatial features of the input of the current layer. The additional $1\times1$ convolutions in the middle branches reduce the number of input channels and the model's complexity while the pooling branch is added in line with traditional CNN networks. The parallel structure with the $1\times1$ convolutions allows networks built on the Inception block to cover a wider range of local features and be stacked in an increasing number of stages and number of units per stage without an uncontrolled blow-up in computational complexity \cite{szegedyGoingDeeperConvolutions2015}. The outputs from the four branches are passed through the Rectified Linear Units (ReLU) \cite{agarapDeepLearningUsing2019} activation function before being concatenated along the channel dimension. It is important to note that, the number of output channels of each convolutional module per layer is a hyperparameter of the Inception block used to control the capacity of the model among the different kernel sizes. For clarity, in this study, we refer to the first $1\times1$ branch as the \textit{transform branch}, the middle branches as \textit{feature branches}, and the last branch as the \textit{pooling branch}.

\begin{figure*}[!ht]
\centering
\includegraphics[width=1.0\linewidth]{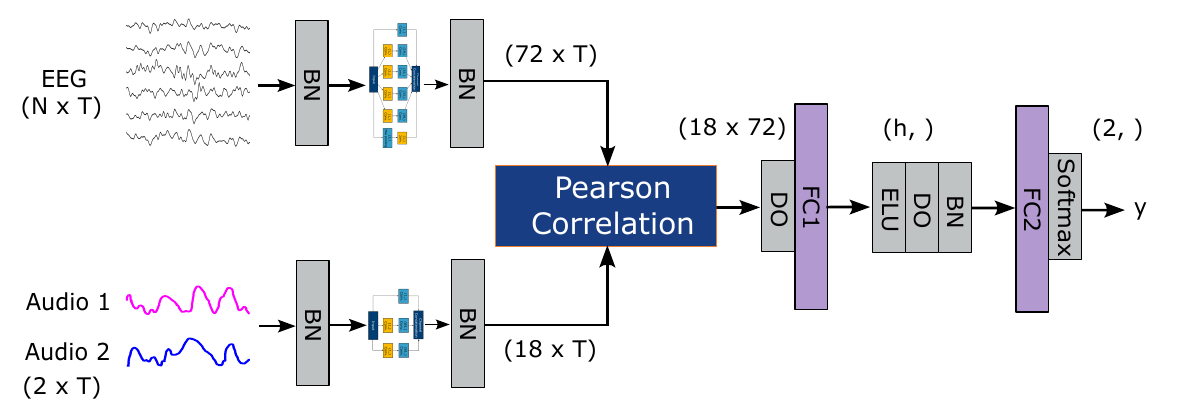}
\caption{The architecture of AADNet. T = length of input data, N = number of input EEG channels, BN = batch normalization, DO = dropout, FC = fully connected, and h = output size of the FC1 layer.}
\label{fig_proposed_E2E}
\end{figure*}

\textbf{AADNet architecture: }
AADNet is a novel envelope-based end-to-end neural network that utilizes the modified Inception block to directly classify the attended speaker without explicitly extracting the audio envelope. The architecture is depicted in \fref{fig_proposed_E2E}. The model comprises two branches: EEG and audio branches. Both preprocessed EEG signals and audio envelopes of the competing speakers sequentially go through a BN layer, a modified Inception block, a $3\times3$ max-pooling layer, and a BN layer. The channel-wise Pearson correlation between the outputs of the two branches is then calculated to extract the relationship between the EEG and audio signals. These correlation values form a feature vector that is flattened out and goes through a DO, an FC layer, and a Softmax activation function to determine the probability of each input audio channel being attended stimulus. It is important to note that, in the audio branch, the competing envelopes are treated separately using the same network Inception block because the two audio channels are independent. The input label was determined based on the index of the attended stimulus in the input audio vector. To prevent the model from being biased by the index of the stimulus, we duplicated each input data, switched the indexes of the stimuli, and changed the input label accordingly to ensure that the attended stimulus was equally distributed.

In this study, the structure of the Inception block was adapted for 1-dimensional data and attention-related features to maximize the AAD performance. Specifically, the Inception blocks in the EEG and audio branches comprised six and four parallel branches, respectively. The kernel sizes of the transform and pooling branches were 1 and 3, as in the original version. All operations used a stride of 1. For EEG signals, the kernel sizes of the four feature branches were set to 19, 25, 33, and 39, covering durations of 0.3, 0.4, 0.5, and 0.6 seconds at a sampling rate of 64 Hz. For audio signals, the kernel sizes of the two feature branches were selected at 65 and 81 corresponding to 1.0 and 1.2 seconds. It is important to note that the input of the audio branch is the audio envelopes which were downsampled to 64 Hz (see the preprocessing step in \sref{sss_preproc_aud}). The pooling branch was empirically omitted since it did not contribute to the overall performance. Details of kernel sizes, and output channels for each module in the Inception block of the EEG and audio branches are shown in \tref{tab2_inception_params}. Here, the kernel sizes were selected based on the potential range that generates the highest correlation between EEG and audio signals in previous studies \cite{osullivanAttentionalSelectionCocktail2015, ciccarelliComparisonTwoTalkerAttention2019a, geirnaertElectroencephalographyBasedAuditoryAttention2021}. The main criterion for selecting the number of branches and output channels per branch was to maximize the number of parallel filters and an appropriate number of parameters relative to the dataset size, such that overfitting was avoided. The model specification in \tref{tab2_inception_params} presents the most successful particular instance tested in our experiments with the two particular datasets described in \sref{ss_dataset}.

\begin{table}[ht]
\caption{\label{tab2_inception_params}Specification of Inception blocks used in AADNet. $Conv.(x, k, y, s)$ represents the convolution with $x$ = input channels, $y$ = output channels, $k$ = kernel size, and $s$ = stride. $N$ is the number of EEG channels, $N_a$ is the number of audio channels, $N_a$ = 1 as the audio streams are treated separately.}
\begin{center}
\resizebox{\linewidth}{!}{
\begin{tabular}{llll}
\hline
\\[-0.2cm]
\textbf{\makecell[l]{Input\\branch}} & \textbf{\makecell[l]{Inception\\branch}} & \textbf{\makecell[l]{Operation}} & \textbf{\makecell[l]{Activation}}\\[0.2cm]
\hline
\\[-0.2cm]
EEG & Transform & Conv.($N$, 1, 32, 1) & ReLU\\[0.3cm]
    & Feature 1 & \makecell[l]{Conv.($N$, 1, 16, 1)\\Conv.(16, 19, 8, 1)} & ReLU \\[0.3cm]
    & Feature 2 & \makecell[l]{Conv.($N$, 1, 8, 1)\\Conv.(8, 25, 8, 1)} & ReLU \\[0.3cm]
    & Feature 3 & \makecell[l]{Conv.($N$, 1, 4, 1)\\Conv.(4, 33, 8, 1)} & ReLU \\[0.3cm]
    & Feature 4 & \makecell[l]{Conv.($N$, 1, 2, 1)\\Conv.(2, 39, 8, 1)} & ReLU \\[0.3cm]
    & Pooling & \makecell[l]{Conv.($N$, 3, $N$, 1)\\Conv.($N$, 1, 8, 1)} & ReLU \\[0.2cm]
\hline
\\[-0.2cm]
Audio & Transform & Conv.($N_a$, 1, 1, 1) & ReLU\\[0.3cm]
    & Feature 1 & \makecell[l]{Conv.($N_a$, 1, 1, 1)\\Conv.(1, 65, 4, 1)} & ReLU \\[0.3cm]
    & Feature 2 & \makecell[l]{Conv.($N_a$, 1, 1, 1)\\Conv.(1, 81, 4, 1)} & ReLU \\[0.2cm]
\hline
\end{tabular}}
\end{center}
\end{table}

\section{Methods}\label{s_methods}

\subsection{Dataset}\label{ss_dataset}
In this work, we use three datasets recorded using a competing-talker setup, which has been used in previous AAD-related studies to address the AAD problem.

\subsubsection{Dataset I - EventAAD}\label{sss_dataset_1}
\textit{Dataset I}, referred to as the EventAAD dataset, was created for investigation of AAD based on cognitive responses to speech events \cite{nguyenCognitiveComponentAuditory2025a}. The data set contains four different paradigms, with a gradual development from sequences of single words towards more and more natural speech situations. The details can be found in the original study \cite{nguyenCognitiveComponentAuditory2025a}. In this study, we only used data from the fourth paradigm, a \squotes{cocktail party} scenario in which participants were simultaneously presented to two competing speech streams in Danish (excerpts from two different audiobooks/radio broadcast news) from two loudspeakers placed one meter from the participant with one placed 60 degrees to the left and the other 60 degrees to the right. In each trial, the participants were instructed to pay attention to one stream while disregarding the other. To sustain the participants' engagement with the task, the participants were probed with a question related to the content of the attended stream after each trial. After the participant responded, feedback was provided to indicate whether the answer was correct. Each subject completed 40 trials, with each trial lasting approximately 30 seconds. EEG was recorded from 32 scalp electrodes and 6 electrodes in each ear using two TMSi Mobita amplifiers with a sampling rate of 1000 Hz. Data was collected from 24 participants. Only the scalp EEG data is used in this study. It should be noted that the stimuli were synthesized using Google Text-to-Speech with the male and female voices randomly selected for both streams in all trials.

\subsubsection{Dataset II - DTU}\label{sss_dataset_2}
\textit{Dataset II} \cite{fuglsangEEGAudioDataset2018} was collected from 18 healthy subjects in a double-walled soundproof booth using a 64-channel BioSemi ActiveTwo system and sampled at a frequency of 512 Hz. The data comprises 60 trials per subject. In each trial, 50-second long competing speech segments, narrated by a male and a female speaker, were binaurally presented to participants through insert earphones in three simulated acoustic conditions: anechoic, mild reverberation, and high reverberation. To spatially separate clean speech signals, the stimuli were simulated using head-related impulse responses for the two speech streams lateralized at ${\pm}60^{\circ}$ along the azimuth direction at a distance of 2.4 meters. Subjects were asked to attend one speaker and ignore the other during each trial. After each trial, they answered a question related to the content of the attended speech stream.

\subsubsection{Dataset III - KUL}\label{sss_dataset_3}
\textit{Dataset III} \cite{dasEffectHeadrelatedFiltering2016} was also collected with a speech competing scenario to address the AAD problem in a soundproof, electromagnetically shielded room using a 64-channel BioSemi ActiveTwo system and sampled at a frequency of 8196 Hz. The data comprises 16 normal hearing subjects with 20 trials per subject. The first 8 trials are 6-minute-long, and the remaining 12 trials, which are repetitions of the first trials, are 2-minute-long. During each trial, two different Dutch stories narrated by male speakers were presented simultaneously through a pair of insert phones. The subjects were instructed to attend to the story in one ear while ignoring the one in the other ear. After each trial, subjects were asked a set of muliple-choice questions about the story. There were two stimulus conditions, i.e., \squotes{dichotic} and head-related transfer filtered with ${\pm}90^{\circ}$ along the azimuth direction. The attended ear and stimulus condition were permuted and kept equally distributed over the trials.

\tref{tab1_datasets} summarizes the details of the three datasets.

\begin{table*}[ht]
\caption{\label{tab1_datasets}Summary of the datasets used in this study.}
\begin{center}
\resizebox{\textwidth}{!}{
\begin{tabular}{lllllll} 
 \hline
 \\[-0.2cm]
 \textbf{\makecell[l]{Dataset}} & \textbf{\makecell[l]{No. of\\subjects}} & \textbf{\makecell[l]{No. of\\trials}} & \textbf{\makecell[l]{Trial\\length (s)}} & \textbf{\makecell[l]{No. of\\channels}} & \textbf{\makecell[l]{Audio presentation}} & \textbf{\makecell[l]{Acoustic condition}}\\[0.2cm]
 \hline
 \\[-0.2cm]
 \makecell[l]{EventAAD} & 24 & 40 & 25 & 32 & \makecell[l]{Loudspeakers,\\${\pm}60^{\circ}$ relative to front direction,\\1 m distance} & \makecell[l]{Shielded with $0.4\,s$ of\\reverberation time} \\ [0.2cm]
 \hline
 \\[-0.2cm]
 \makecell[l]{DTU} & 18 & 60 & 50 & 64 & \makecell[l]{Insert earphones,\\ simulated speaker ${\pm}60^{\circ}$ direction\\using a head-related transfer function,\\2.4 m distance} & \makecell[l]{Anechoic, mildly,\\and highly reverberant} \\[0.2cm]
 \hline
  \\[-0.2cm]
 \makecell[l]{KUL} & 16 & 8+4 & 360/120 & 64 & \makecell[l]{Insert earphones,\\ simulated speaker ${\pm}90^{\circ}$ direction\\using a head-related transfer function\\ and dichotic} & \makecell[l]{Electromagnetically\\ shielded room} \\[0.2cm]
 
 \hline
\end{tabular}}
\end{center}
\end{table*}

\subsection{Data pre-processing}\label{ss_preprocess}
In this section, we present the processing steps that were applied on the three presented datasets for the LSR, CCA, and the proposed AADNet models. For the NSR model, we followed the processing steps proposed in the original study \cite{thorntonRobustDecodingSpeech2022}.

\subsubsection{EEG}\label{sss_preproc_eeg}
The EEG data were first band-pass filtered using a zero-phase FIR filter with cutoffs at 0.5 Hz and 32 Hz to eliminate slow drift and irrelevant high frequencies. This frequency band was also applied in various studies \cite{detaillezMachineLearningDecoding2020, ciccarelliComparisonTwoTalkerAttention2019a, geirnaertElectroencephalographyBasedAuditoryAttention2021} and is reported to be relevant for envelope-based AAD studies. The data for each channel were then downsampled to 64 Hz, re-referenced to the average of all channels (32 channels for the EventAAD dataset and 64 channels for the DTU and KUL dataset), and zero-centered. For the NSR model, the EEG data were band-pass filtered from 0.25 to 36 Hz using a Hamming window, FIR filter, resampled to 125 Hz, and standardized to have zero mean and unit variance. The pre-processing pipeline was implemented using the Python MNE package version 1.2.0 \cite{gramfortMEGEEGData2013} and SciPy package version 1.10.1 \cite{virtanenSciPyFundamentalAlgorithms2020}.

\subsubsection{Audio}\label{sss_preproc_aud}
The audio envelopes were extracted using the compressed subband envelopes, resembling the processing of speech signals in the human auditory system \cite{biesmansAuditoryInspiredSpeechEnvelope2017b}. Specifically, we first applied a gammatone filter bank with an equivalent rectangular bandwidth (ERB) equal to 1.5 Hz, and center frequencies ranging from 150 Hz to 4 kHz. The compressed envelope of each subband was computed with a power law exponent of 0.6. The final envelope was then obtained by summing all the subband envelopes. For the NSR model, the envelopes were extracted by taking the absolute value of the Hilbert transform of each speech stream. The envelopes were then low-pass filtered with a cut-off frequency of 50 Hz using an FIR filter, Hamming window, 12.5 Hz transition bandwidth, and resampled to 125 Hz. Both envelope extraction methods were implemented in Python using the SciPy package.

\subsection{Evaluation procedure}\label{ss_training}
As pointed out in Rotaru \etal \cite{rotaruWhatAreWe2024}, the DL approach is susceptible to capturing subtle biases, such as within-trial fingerprints of neural activities, even across different subjects. This may lead to artificially high decoding accuracies. To mitigate this potential bias, it is essential to perform appropriate data splitting into training, validation, and test sets. In this study, we carefully employ a trial-based cross validation to evaluate the SS and SI models for all investigated methods. In the remainder of this paper, we refer to the length of data used by the models to make a decision as the analysis window length.

\subsubsection{SS models}\label{sss_ss_model}
Each subject's data was divided into eight folds on a trial basis. Seven of the eight folds were split into a training set and a validation set in a 4:1 ratio (also on a trial basis) while the remaining fold was used as the test set. The data of each trial were further segmented into smaller windows for training and testing, except when training LSR, CCA, and NSR models, where the training data were concatenated across all training trials. The models were trained using the training and validation sets and evaluated on the test set to obtain the performance for each fold. This procedure was repeated for eight folds. The accuracy of each SS model was calculated as the average accuracy across the eight folds.

\subsubsection{SI models}\label{sss_si_model}
\begin{figure}[!ht]
\centering
\includegraphics[width=1.0\linewidth]{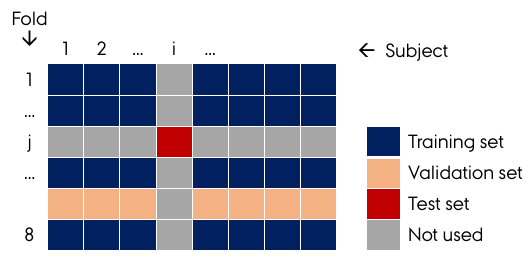}
\caption{Visual illustration of the cross-trial leave-one-subject-out (LOSO) cross-validation scheme. The data is split for training at fold $j$ when holding out subject $i$ for testing.}
\label{fig_cross_validation}
\end{figure}

For the SI model, data from one subject were held out for the test set while the data from the remaining subjects were used for training and validation. However, the EventAAD dataset was collected using the same pairs and order of stimuli for all subjects which could lead to potential leakage of stimuli information from the training set to the test set, allowing the model to indirectly learn to perform the AAD task. This is not the case for the DTU and KUL datasets where the order and pair of stimuli were randomly mixed. To ensure that the attended stimuli in the testing data were not presented as the attended stimuli in the training and validation data, we performed an eight-fold leave-one-subject-out (LOSO) cross-validation scheme. The data from the test subject were divided into eight folds. Only one fold was held out for the test set while the other seven folds were not used in the current iteration. For all the remaining training subjects, the trials whose attended stimuli being in the test set were left out of the training set, and the remaining trials were pooled together across subjects and split into a training set and a validation set in a 4:1 ratio (based on a trial basis). Models were trained using the training and validation set and evaluated on the test set to obtain a fold performance. In this manner, all the attended stimuli used for the test subject were not used as the attended stimuli during training, even for the other subjects. See \fref{fig_cross_validation} for a visual illustration of the procedure. This procedure was repeated across eight folds. The accuracy of each SI model was calculated as the average accuracy across the eight folds.

\subsection{Performance metrics}\label{ss_metrics}
We used two metrics to quantify the performance of the investigated models.
\subsubsection{Classification accuracy}\label{ss_metrics_acc}
Classification accuracy was assessed as a function of the analysis window length. For each window length, the test data were transformed into a data matrix with an overlap of at least 50\%. Each stimulus-reconstruction model (LSR and NSR) predicted the attended audio using the specified window length of EEG data. The reconstructed envelope was then compared with the attended and unattended envelopes using the Pearson correlation. An attempt was correct if the correlation score with the attended envelope was higher than that of the unattended envelope. For the CCA method, the model attempted to estimate the two correlation vectors between EEG data and the attended and unattended envelopes. The difference between the attended and the unattended correlation vectors was passed through the binary LDA classifier. The attempt was correct if the output of LDA was 1 and vice versa. For the AADNet, the output was considered correct if it matched the input label. Accuracy was calculated as the ratio of the number of correct attempts and the number of input data.

\subsubsection{Minimal expected switch duration}\label{ss_metrics_mesd}
The accuracy metric described above is a window length-dependent quantification. It is expected that the longer the window length, i.e., the more information available, the higher the accuracy. To obtain a more effective performance measurement, we also calculated the minimal expected switch duration (MESD), an interpretable performance metric for AAD algorithms in the context of neuro-steered gain control \cite{geirnaertInterpretablePerformanceMetric2020}. The MESD addresses the trade-off between AAD accuracy and decision time by modeling an adaptive gain control system in a hearing device as a Markov chain, and based on that calculating the minimal expected time required to switch the operation mode after an attention switch of the user. A lower MESD corresponds to better AAD performance and vice versa. In this study, the MESD was calculated using the Python MESD toolbox \cite{geirnaertMESDToolbox2019}.

\subsection{Hyperparameter choice and model training} \label{ss_implementations}
The hyper-parameters of the linear methods were selected based on the recommended values used in the original studies as discussed in \sref{s_algs}. To determine the optimal $J$ value in the CCA method, we first performed inner cross-validation on the training data for the LDA model to find the optimal $J_f$ for each fold and each analysis window from the outer cross-validation of the CCA model. The final $J$ values for SS and SI models were obtained by taking the minimum value across analysis windows of the average and grand average of the $J_f$, respectively. These final $J$ values were then used to validate the corresponding CCA models again on the test data to obtain the final performances. For the NSR model, we performed full hyperparameter optimizations for each dataset based on the guidelines from the original study \cite{thorntonRobustDecodingSpeech2022}.

The proposed AADNet was trained with the Cross-Entropy loss function using the AdamW optimizer \cite{loshchilovDecoupledWeightDecay2019} which is advantageous in decoupling L2 regularization, i.e., weight decay, from the learning rate so that we have less number of hyper-parameters to search. The hyperparameters were chosen via a random search within configurations on the EventAAD dataset as follows: batch size = (32, 64, 128), weight decay = ($10^{-4}$, $10^{-3}$, $10^{-2}$, $10^{-1}$), dropout = (0.5, 0.4, 0.3, 0.2, 0.1) and number of output channel of FC1 layer, i.e., hidden units, $h$ = (32, 16, 0). In the case of $h$ = 0, the ELU, DO, BN, and FC2 after the FC1 layer were omitted. The learning rate was fixed at $5\times10^{-5}$. The optimal hyperparameters were then fixed for the other two datasets. During the training process, the model was saved if the validation loss decreased and the training process was stopped if the validation loss did not decrease in at least 1 out of 5 consecutive epochs. Due to limited data sets for SS models that potentially lead to overfitting, for the NSR and the proposed AADNet models, we started training the SI models first and used the saved model to fine-tune the SS models for the subject that was left out with smaller tuned learning rates.

The LSR cross-validation method was implemented using the Scikit-learn package \cite{pedregosaScikitlearnMachineLearning2011} with \code{linear\_model.RidgeCV} function while the CCA method uses \code{cross\_decomposition.CCA} and \code{discriminant\_analysis.LinearDiscriminant-} \code{Analysis} functions. The NSR and AADNet were implemented using the Pytorch framework \cite{paszkePyTorchImperativeStyle2019a}.

\section{Results}\label{s_results}
\subsection{SS models}\label{ss_results_SS}
\begin{figure*}[!ht]
\centering
    \subfigure[]
    {
     \includegraphics[width=0.48\linewidth]{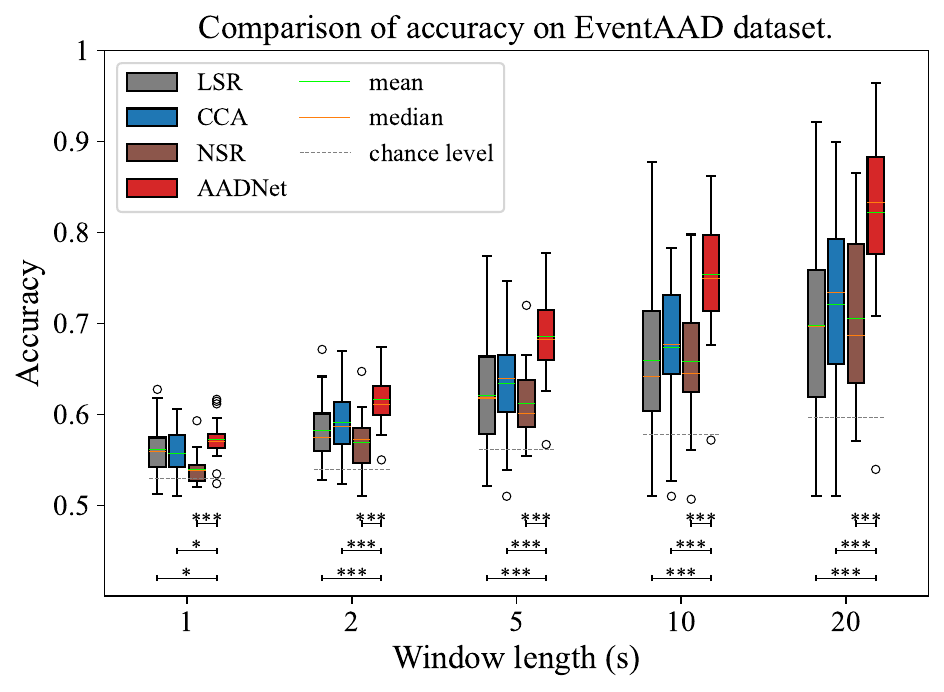}
     \label{fig_EventAAD_SS_acc}
    }
    \subfigure[]
    {
     \includegraphics[width=0.42\linewidth]{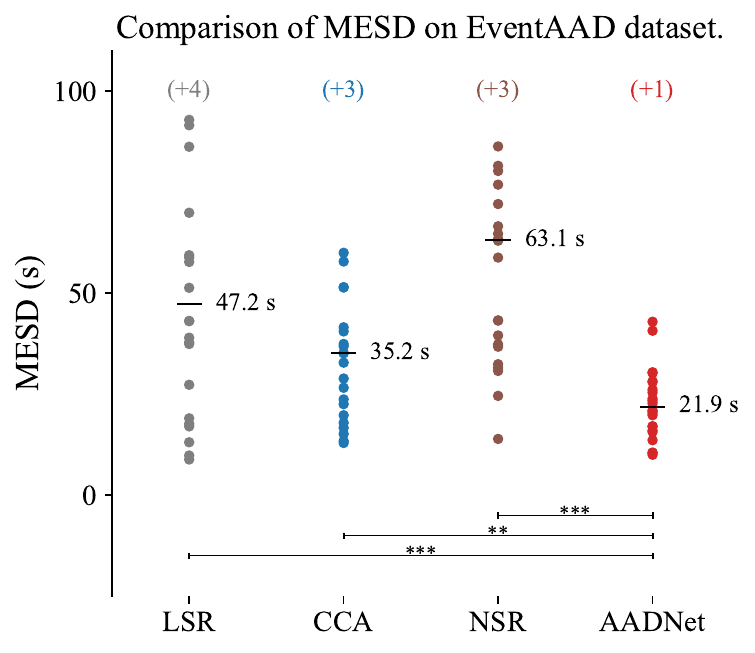}
     \label{fig_EventAAD_SS_mesd}
    }
    \\
    \subfigure[]
    {
     \includegraphics[width=0.48\linewidth]{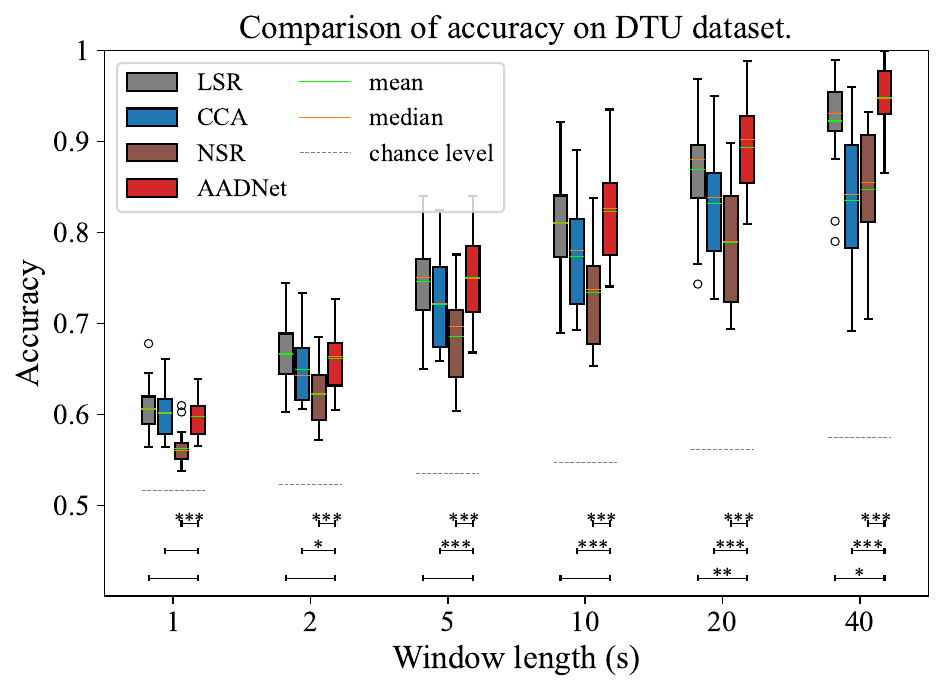}
     \label{fig_DTU_SS_acc}
    }
    \subfigure[]
    {
     \includegraphics[width=0.42\linewidth]{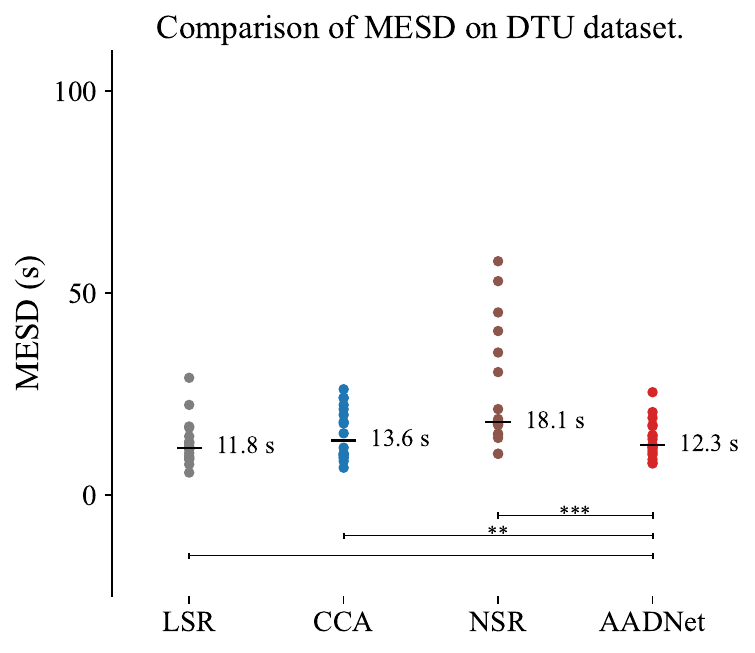}
     \label{fig_DTU_SS_mesd}
    }
    \\
    \subfigure[]
    {
     \includegraphics[width=0.48\linewidth]{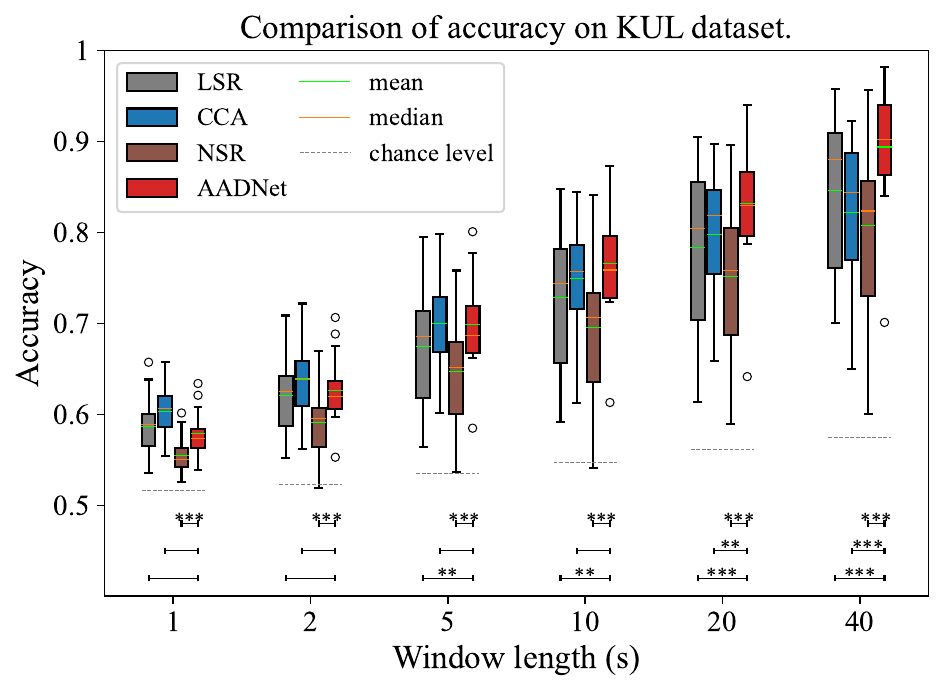}
     \label{fig_KULeuven_SS_acc}
    }
    \subfigure[]
    {
     \includegraphics[width=0.42\linewidth]{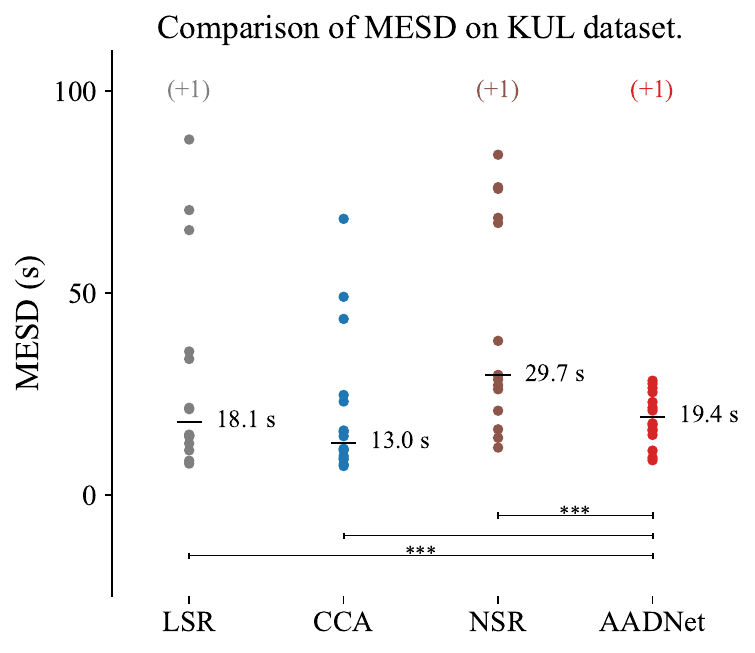}
     \label{fig_KULeuven_SS_mesd}
    }    
\caption{Comparison of the SS models on the three datasets. (a), (b) The accuracies and MESD of the four methods on the EventAAD dataset. (c), (d) The accuracies and MESD of the four methods on the DTU dataset. (e), (f) The accuracies and MESD of the four methods on the KUL dataset. The number of data points with an MESD of $>100\,$s is indicated as (+x) and was included in the computation of the medians. Statistical significance is denoted by asterisks (None: $p \ge 0.05$; *: $0.05 > p \ge 0.01$; **: $0.01 > p \ge 0.001$; ***: $0.001 > p$.}
\label{fig_results_SS}
\end{figure*}

Decoding accuracies of the SS models for each method on the EventAAD, DTU and KUL datasets are shown in \fref{fig_EventAAD_SS_acc}, \fref{fig_DTU_SS_acc}, and \fref{fig_KULeuven_SS_acc}, respectively. The box plots represent the accuracy distribution across 24 subjects for the EventAAD dataset, 18 subjects for the DTU dataset, and 16 subjects for the KUL dataset. The chance performance was computed as the $95^{th}$ percentile point of a binomial distribution with $p = 0.5$ and $n$ equal to the number of non-overlapping windows in the test set. We compared the proposed AADNet with other baseline methods and tested the significance using the paired permutation test \cite{marisNonparametricStatisticalTesting2007} with the Bonferroni correction. The test results are shown at the bottom of the figures. On the EventAAD dataset, the mean accuracy of AADNet increases from 0.572 at 1$\,$s to 0.822 at 20$\,$s. For the DTU and KUL datasets, the mean accuracies at 1$\,$s are 0.597 and 0.579, respectively, reaching 0.949 and 0.894, at 40$\,$s. Notably, AADNet significantly outperforms all baseline methods across all window lengths on the EventAAD dataset ($p\,<\,0.05$). For the DTU and KUL datasets, AADNet still performs significantly better than the other in the majority of comparisons. However, it does not surpass the LSR method for short windows on the DTU dataset ($\leq\,$10$\,$s) and the CCA method on the KUL dataset ($\leq\,$10$\,$s). The results also demonstrated that the NSR method is the lowest-performing model in all three datasets.

Additionally, we calculated the per-subject MESD values for each model on the three datasets and presented them in \fref{fig_EventAAD_SS_mesd}, \fref{fig_DTU_SS_mesd}, and \fref{fig_KULeuven_SS_mesd}, respectively. Significant differences in the median values between the proposed model and the others were also tested using paired permutation tests.
We found that the proposed AADNet achieved a significantly lower MESD value of 21.9$\,$s on the EventAAD dataset. It also achieved 12.3$\,$s and 19.4$\,$s on the DTU and KUL datasets, respectively. In comparison, the NSR model, consistent with the lowest performance in accuracy, achieved MEDS values of 63.1$\,$s, 18.1$\,$s, and 29.7$\,$s for the EventAAD, DTU, and KUL datasets, respectively.

\subsection{SI models}\label{ss_results_SI}
\begin{figure*}[!ht]
\centering
    \subfigure[]
    {
     \includegraphics[width=0.48\linewidth]{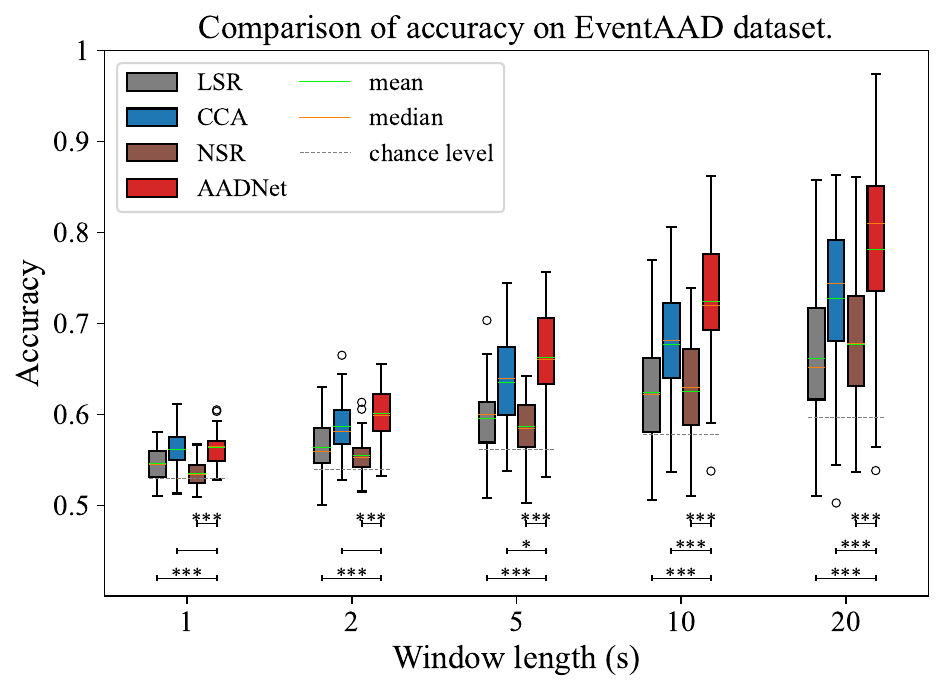}
     \label{fig_EventAAD_SI_acc}
    }
    \subfigure[]
    {
     \includegraphics[width=0.42\linewidth]{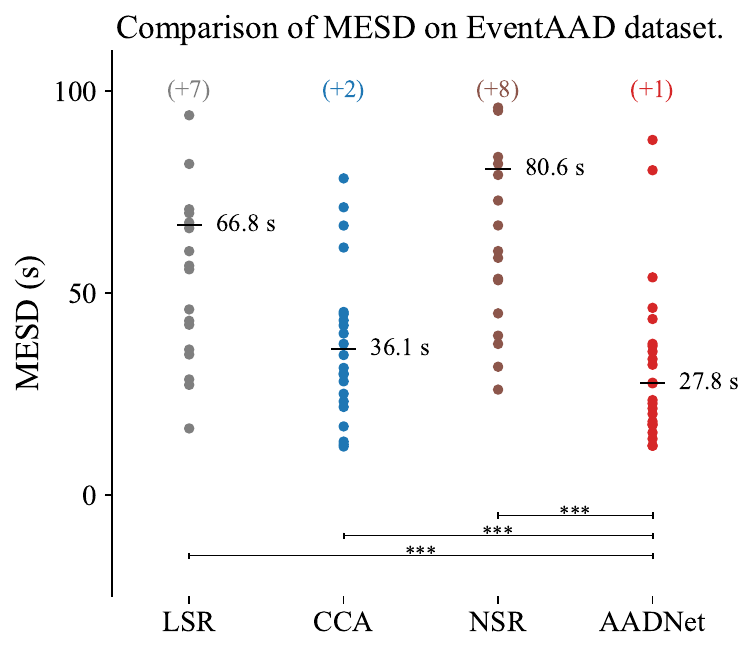}
     \label{fig_EventAAD_SI_mesd}
    }
    \\
    \subfigure[]
    {
     \includegraphics[width=0.48\linewidth]{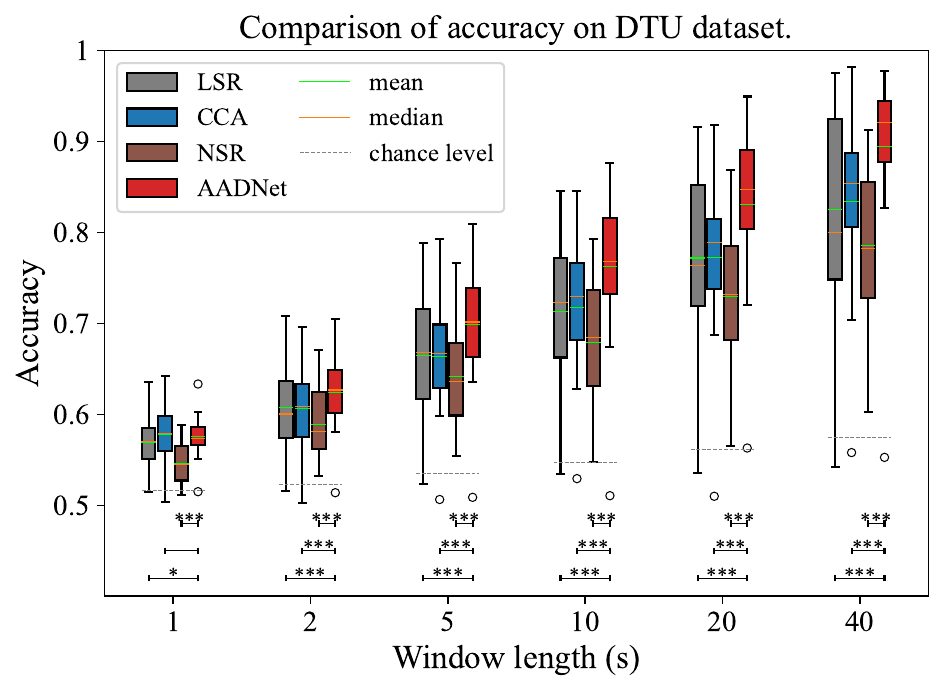}
     \label{fig_DTU_SI_acc}
    }
    \subfigure[]
    {
     \includegraphics[width=0.42\linewidth]{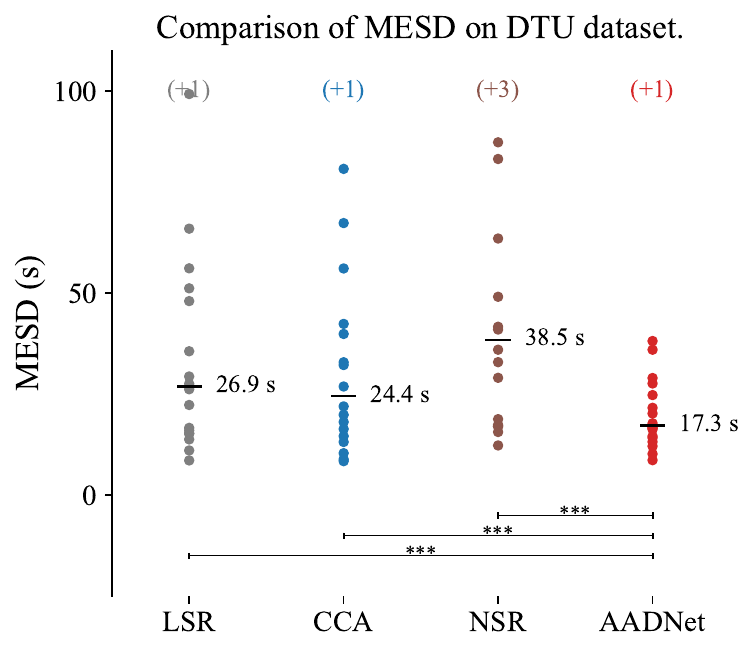}
     \label{fig_DTU_SI_mesd}
    }    
    \\
    \subfigure[]
    {
     \includegraphics[width=0.48\linewidth]{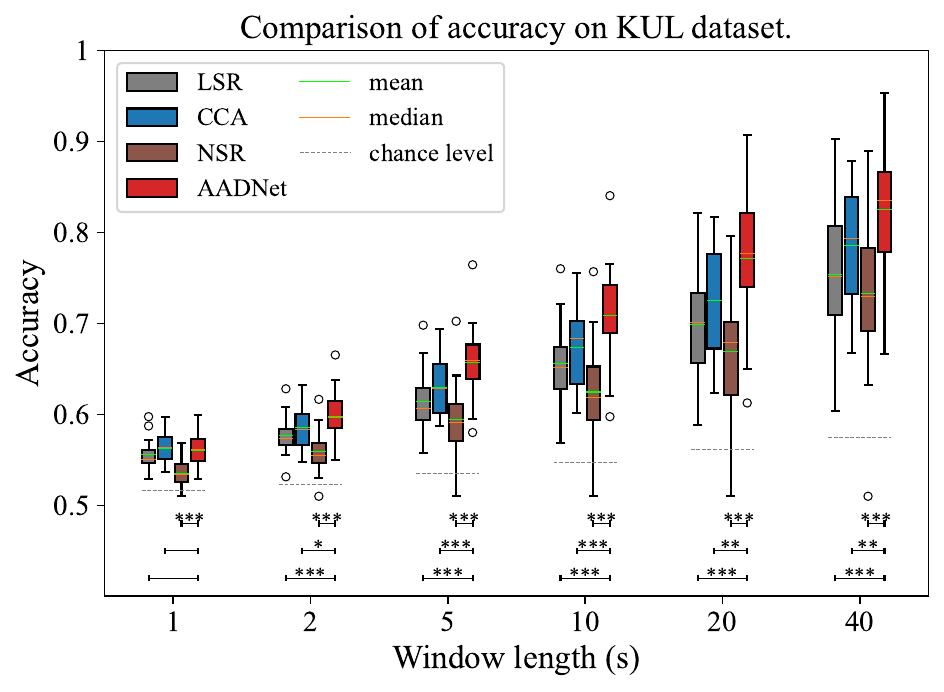}
     \label{fig_KULeuven_SI_acc}
    }
    \subfigure[]
    {
     \includegraphics[width=0.42\linewidth]{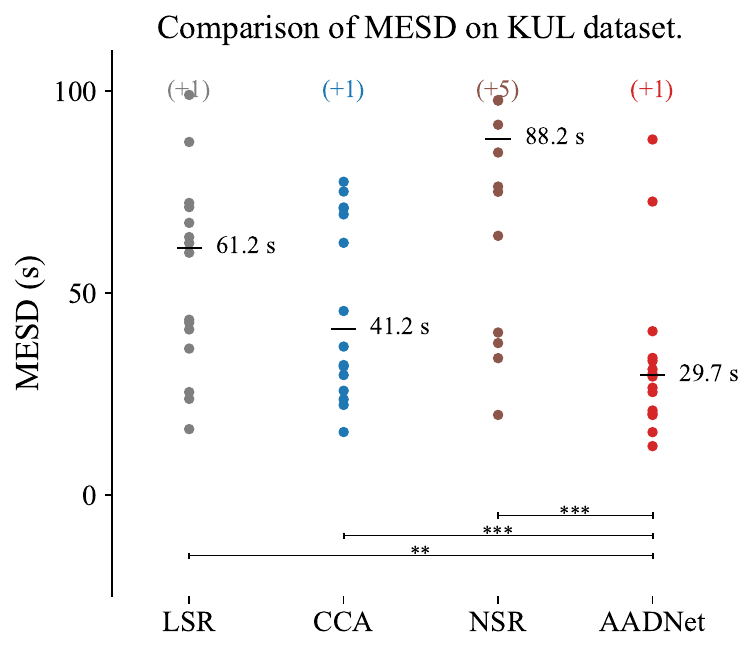}
     \label{fig_KULeuven_SI_mesd}
    }     
\caption{Comparison of the SI models on the three datasets. (a), (b) The accuracies and MESD of the four methods on the EventAAD dataset. (c), (d) The accuracies and MESD of the four methods on the DTU dataset. (e), (f) The accuracies and MESD of the four methods on the KUL dataset.}
\label{fig_results_SI}
\end{figure*}

To evaluate how well AADNet generalizes to new subjects, we performed the SI validation scheme described in \sref{sss_si_model}. The results of SI models are shown in \fref{fig_results_SI}. Similar to the SS results, \fref{fig_EventAAD_SI_acc}, \fref{fig_DTU_SI_acc}, and  \fref{fig_KULeuven_SI_acc} present the test accuracies across subjects while \fref{fig_EventAAD_SI_mesd}, \fref{fig_DTU_SI_mesd}, and \fref{fig_KULeuven_SI_mesd} present the MESDs. Generally, the SI models obtained a lower performance compared to the corresponding SS models (see \tref{tab3_ss-si}). The results demonstrate that AADNet significantly outperforms the other methods for almost all window lengths on all three datasets, except the CCA model at 1$\,$s and the LSR model at 1$\,$s (only on the KUL dataset). At this analysis window, AADNet and CCA models achieved similar performance, with accuracies of 0.563 and 0.561 for EventAAD, 0.575 and 0.577 for DTU, and 0.56 and 0.563 for KUL. The mean accuracies of AADNet reach 0.781 at 20$\,$s, 0.894 at 40$\,$s, and 0.826 at 40$\,$s, for EventAAD, DTU, and KUL datasets, respectively. Regarding MESD, AADNet achieves significantly lower values of 27.8$\,$s, 17.3$\,$s, and 29.7$\,$s on the EventAAD, DTU, and KUL datasets, respectively. The NSR method again performed the worst in both metrics.

\begin{table}[!ht]
\caption{Performance drop (in percentage points of accuracy) of SI models compared to corresponding SS models.}\label{tab3_ss-si}
\centering
\resizebox{1.0\linewidth}{!}{
\begin{tabular}{|c|c|c|c|c|c|c|c|}
\hline
\multirow{2}{*}{Dataset} & \multirow{2}{*}{Models} & \multicolumn{6}{c|}{Window lengths (s)}\\
\cline{3-8}
    &   & 1 & 2 & 5 & 10 & 20 & 40\\
\hline
\multirow{4}{*}{EventAAD} & LSR & 1.6 & 1.8 & 2.5 & 3.5 & 3.7 &\\
                       & CCA & -0.3 & 0.5 & -0.2 & -0.4 & -0.7 &\\
                       & NSR & 0.5 & 1.4 & 2.5 & 3.4 & 2.9 &\\
%                       & NSR & 0.5 & 1.4 & 2.5 & 3.4 & 2.9 &\\
                       & AADNet & 0.9 & 1.5 & 2.3 & 3.0 & 4.1 &\\   
\hline
\multirow{4}{*}{DTU}      & LSR & 3.7 & 5.9 & 8.1 & 9.8 & 9.7 & 9.8   \\
                            & CCA & 2.5 & 4.2 & 5.7 & 5.6 & 6.0 & 0.0   \\
                            & NSR & 1.7 & 3.3 & 4.4 & 5.5 & 6.0 & 6.1   \\
                            %& NSR & 1.7 & 3.3 & 4.4 & 5.5 & 6.0 & 6.1   \\
                            & AADNet & 2.2 & 3.6 & 5.1 & 6.1 & 6.4 & 5.5  \\                             
\hline 
\multirow{4}{*}{KUL}   & LSR & 3.0 & 4.3 & 6.0 & 7.3 & 8.4 & 9.3   \\
                            & CCA & 4.0 & 5.3 & 6.9 & 7.6 & 7.2 & 3.6   \\
                            & NSR & 2.0 & 3.2 & 5.3 & 7.0 & 8.3 & 7.5   \\
%                            & NSR & 2.0 & 3.2 & 5.3 & 7.0 & 8.3 & 7.5   \\
                            & AADNet & 1.9 & 2.9 & 4.2 & 5.7 & 6.1 & 6.8  \\                             
\hline
\end{tabular}}
\end{table}

\section{Discussion and Conclusion}\label{s_discussion}
\subsection{Classification performance}\label{ss_discuss_performance}
In line with most AAD studies, we started by training SS models and evaluating them using a multi-fold cross-validation. However, we found unexpectedly low performances of the NSR and AADNet models. While the exact reason for this is unclear, we conjecture that it is due to the higher number of parameters in these models. In consequence, the training requires a larger amount of data and the models are more susceptible to overfitting. This is particularly the case for the SS models and less of a problem for the SI models. To address this challenge, the SS models were trained by finetuning the pretrained SI model, achieving promising results. As shown in \fref{fig_results_SS}, the proposed model outperformed the baseline methods on the EventAAD dataset. Although it did not show superiority over the LSR method on the DTU dataset and the CCA method on the KUL dataset, AADNet achieved the best accuracy at long analysis windows($>$ 20$\,$s) and an overall competitive performance on these two datasets, showcasing its strong adaptability across varying data distributions. This performance advantage could be attributed to two main factors: the nonlinearity and parallelized spatiotemporal convolutions, which allow the model to more effectively learn the audio representation in the human auditory system and capture the relationship between the audio stimulus and the brain signals.

We also trained and evaluated SI models using the cross-trial LOSO validation scheme. Although all models yielded accuracies significantly above the chance performance, there was a consistent drop in performance compared to the corresponding SS models (see \tref{tab3_ss-si}). This difference is somewhat expected, as the SS models, unlike the SI models, are fine-tuned to capture the unique characteristics of individual subjects. The results in \fref{fig_results_SI} showed a superior performance of the proposed AADNet compared to the other methods with gaps of 5.4, 5.9, and 4.0 of percentage points at the longest window length to the best baseline method on the EventAAD, DTU, and KUL datasets, respectively. This demonstrated a better generalization for new subjects. We conjecture that the multiple parallelized convolutions in the Inception structure, which provide extensive coverage of the spatiotemporal dependency between the speech envelope and the brain response across subjects, play a key role in this enhanced performance. Even though the improvement may seem modest, it could have a valuable contribution to the AAD field due to the training-free advantage of the SI model for the new subjects and the use of fixed hyperparameter sets across datasets. This advantage makes the SI model more feasible to be integrated into real-life applications.

In regard to MESD, we showed that the proposed AADNet consistently and significantly achieved the lowest MESD values among SI models across all three datasets. Since MESD represents the minimal expected time for an AAD-based gain control system to switch operation modes, these improvements demonstrated that the proposed AADNet holds promise for integrating the AAD algorithm into a real-time gain control system for hearing-assistive devices. However, due to its moderate performance at short window lengths, the proposed SS model did not outperform the linear methods in MESD metrics. This motivates future work to develop a more advanced model to improve the performance at short window lengths.

%It is important to note that this study found the NSR method to perform modestly compared to the LSR method. This contrasts with the original NSR study \cite{thorntonRobustDecodingSpeech2022}, which reported that NSR outperformed LSR. A possible explanation is that the NSR study, along with other deep learning-based methods, employed less rigorous validation procedures that did not fully exclude training data, including attended training stimuli and/or trials, from the test set, potentially resulting in overly optimistic decoding accuracies. This highlights the need for proper cross-validation procedures to reliably assess model performance in the AAD field.

\subsection{Channel distribution of the AAD performance} \label{ss_channels_distribution}
To examine the importance of each channel to the overall performance of the AADNet, a leave-one-channel-out analysis was performed on the SI models for the three datasets. For each dataset, we trained and validated $N$ models in which one channel was left out at a time and the EEG data was referenced to the average of the remaining $N-1$ channels, where $N$ is the number of EEG electrodes used in that dataset. The performance drop of each model, compared to the original SI model, was then calculated and used to obtain the heatmaps (see the supplementary materials). The heatmaps consistently show a peak at the area of electrode C5 across the three datasets, indicating an active brain region involved in the speech attention task. This result is reasonable, as this area is close to the region of the primary auditory cortex and the superior temporal gyrus, which are primarily involved in speech comprehension and the envelope tracking process \cite{kubanekTrackingSpeechEnvelope2013,keitelSharedModalityspecificBrain2020}. Furthermore, the left lateralization of the heatmaps aligns with previous findings that language processing is strongly lateralized to the left hemisphere \cite{binderHumanBrainLanguage1997}.

\subsection{Limitations}\label{ss_discuss_limitations}
In this study, only four approaches were considered: a linear (LSR) and a non-linear (NSR) backward model, a forward-backward combined model (CCA), and the proposed direct classification model. We did not test the forward approach as it is an underperforming method \cite{wongComparisonRegularizationMethods2018a, alickovicTutorialAuditoryAttention2019b, geirnaertElectroencephalographyBasedAuditoryAttention2021}. Moreover, there have been several studies using neural networks and different features to solve different attention-decoding tasks, including SpkI \cite{mirkovicDecodingAttendedSpeech2015, dasEffectHeadrelatedFiltering2016, detaillezMachineLearningDecoding2020, ciccarelliComparisonTwoTalkerAttention2019a, thorntonRobustDecodingSpeech2022} and LoA \cite{vandecappelleEEGbasedDetectionLocus2021a, geirnaertFastEEGBasedDecoding2021, suSTAnetSpatiotemporalAttention2022a, tanveerDeepLearningbasedAuditory2024}. It is challenging to make a direct comparison across these studies due to variations in datasets, used features, tasks, and analysis window lengths used to report results. Here, only the methods in the SpkI task that exploit the envelope-following response are included. For the sake of completeness and transparency, it must be mentioned that we also implemented the methods proposed by De Taillez \etal \cite{detaillezMachineLearningDecoding2020}, Ciccarelli \etal \cite{ciccarelliComparisonTwoTalkerAttention2019a}, and Kuruvila \etal \cite{kuruvilaExtractingAuditoryAttention2021}. However, despite our best efforts in implementation and validation, these methods performed significantly worse than others, so their results are not included here. A similar observation was reported in \cite{geirnaertElectroencephalographyBasedAuditoryAttention2021}.

\subsection{Deep-learning methodology for direct AAD}\label{ss_discuss_DL}
This study aimed to enhance AAD by proposing a deep learning model that directly classifies the attended audio stimulus. We developed an end-to-end neural network to address the AAD problem and achieved a significant improvement in performance compared to other baseline methods. In the remainder of this section, we discuss the advantages and disadvantages of the proposed AADNet as well as the DL approach to consider in future work to leverage the AAD performance.

The architecture of the network was inspired by the Inception block, which comprises multiple convolutional branches. The convolution in each branch acts as a spatiotemporal filter to capture how the speech is encoded in the neural response. The kernel size plays a role in limiting this relationship with a specific time delay. This structure allows the model to combine information in a way similar to the CCA method. However, in the CCA method, the optimal filter length may not work well for a wide range of subjects and datasets. With the parallel architecture, the model can be extended by adding additional branches with different kernel sizes. This may enhance generalization performance, provided that a sufficient amount and diversity of training data are available.

The convolutional kernel size of 1 also plays an important role in constructing the network. In the transform branch, it helps transfer features from the previous layers and bypass the current layer if the features in that layer are not relevant. This allows us to construct a deeper neural network while maintaining control over the model's complexity. This feature is crucial for improving the capability of models that address problems involving small datasets like AAD or other EEG-based applications. Additionally, the kernel size of 1 in the feature layer also plays a role in channel selection to reduce irrelevant information and therefore somewhat prevents the model from overfitting, a common issue in BCI applications.

Another important design of the network in this study is the multi-audio input. This allows the network, at any time point during the training phase, to have access to all audio streams and have a higher degree of freedom to pick relevant features to maximize the difference between audio streams. This is different from the previous AAD studies where the models were forced to find the relationship between the brain response and the attended/unattended stimulus. Additionally, this structure allows us to easily augment the data by shuffling the order of audio stimuli. This data augmentation, in turn, could eliminate the potential biases in directional attention, which are inevitable in some datasets, as pointed out in \cite{rotaruWhatAreWe2024}.

A common challenge with DL models is that they require significantly more data to achieve good performance compared to less complex models. This is indeed also a challenge in AAD, where data sets are very limited, and collecting additional data is very resource-intensive. However, ongoing advancements in DL methods, such as more sophisticated architectures like the Inception structure and improved training methodologies in transfer learning and regularization techniques, are progressively alleviating the issue of limited data.

\section*{Acknowledgment}
This work was funded by the William Demant Foundation, grant numbers 20-2673, and supported by Center for Ear-EEG, Department of Electrical and Computer Engineering, Aarhus University, Denmark and a junior postdoctoral fellowship fundamental research (for S. Geirnaert, grant No. 1242524N) and a travel grant for a long stay abroad (for S. Geirnaert, grant No. V418923N) from the Research Foundation Flanders (FWO).

\bibliography{references}
\end{document}